\documentclass[jkps,preprint,fleqn,showpacs,showkeys]{revtex4}
\usepackage{graphicx}
\usepackage{amssymb}
\usepackage{amsmath}
\usepackage{bm}
\begin{document}
\setcounter{page}{1}
\title[]{Inclusive production of $W$ Boson in $pp$ Collisions in a range of center-of-mass energy from 7 to 100 TeV}
\author{Hasan \surname{Ogul}}
\email{hogul@sinop.edu.tr}
\thanks{Fax: +90 368 271 4152}
\affiliation{Department of Nuclear Engineering, Sinop University, Sinop, 57000, TURKEY}
\author{Emrah \surname{Tiras}}
\affiliation{Department of Physics and Astronomy, Iowa State University, Ames, IA 50011, USA}
\author{Kamuran \surname{Dilsiz}}
\affiliation{Department of Physics, Bingol University, Bingol 12000, Turkey}
\date[]{Received: xxx, 2018}

\begin{abstract}
The proton-proton collision energy at Large Hadron Collider (LHC) has been 7, 8 and 13 TeV recently with the goal of reaching to 14 TeV which is the maximum capacity of the LHC. However,  there is still more physics yet to be explored and tested beyond the energy regime of the LHC to reach new discoveries. Therefore, a new collider bigger than the LHC machine, which will be able to collide protons at 100 TeV center-of-mass energy, is under consideration by the high-energy physics community. To provide an insight to the transition from LHC to 100 TeV collider, some properties of W are investigated in a range of collision energy from 7 to 100 TeV using HERAPDF2.0, MMHT2014, NNPDF3.1 and CT14 NNLO PDF models at NNLO QCD. The considered properties are the production rate of W boson, the change of uncertainties (PDF, renormalization and factorization scales, strong coupling constant, model and parameterization), and W boson lepton charge asymmetry.
\end{abstract}



\maketitle

\section{Introduction}
\label{wboson}
Inclusive production cross sections of $W^{\pm}$ bosons times leptonic branching ratios are calculated using FEWZ 3.1 MC generator program \cite{FEWZ}. The reliability of this program and its comparison with other MC generators were investigated and found at very good level \cite{hasanQCD,hasanZboson,chargeasymmetry7,chargeasymmetry8}.\par
The total and fiducial cross section predictions are performed for born level (pre-QED FSR) leptons based on HERAPDF2.0 \cite{herapdf20}, MMHT2014 \cite{mmht2014}, NNPDF3.1 \cite{nnpdf31} and CT14 \cite{ct14pdf} NNLO PDF models at NNLO QCD. The defined fiducial region to predict the rates of $W^{\pm}$ bosons is that the leptons are required to have $|\eta| <$ 2.4 and $p_{T} >$ 25 GeV.\par
\begin{figure}[h!]
\begin{center}
\includegraphics[width=0.45\textwidth,height=0.2\textheight]{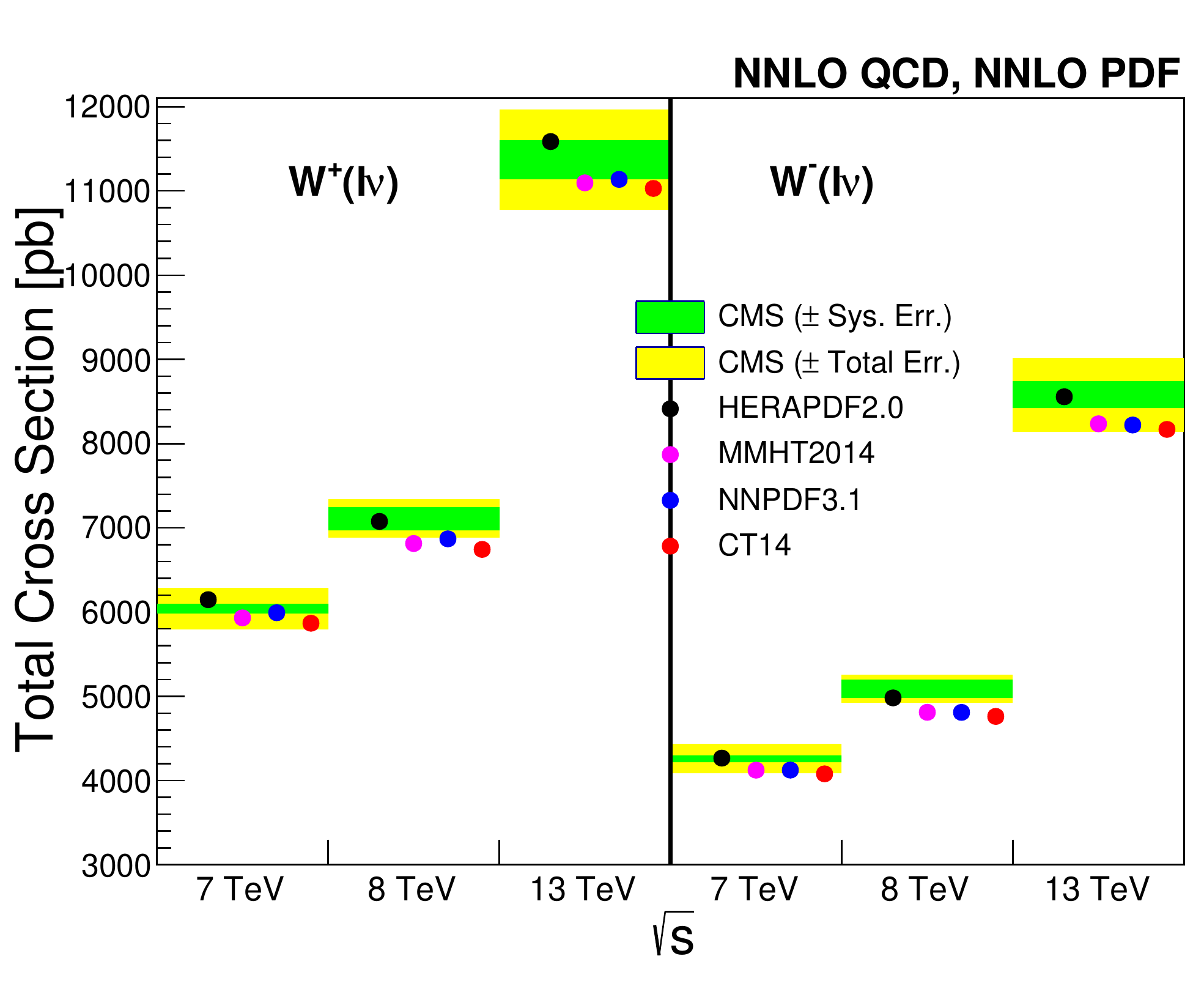}
\caption{Predicted and measured values of $\sigma_{W^{+}}^{Total}$.BR($W^{+}\rightarrow l^{+}\nu)$ and $\sigma_{W^{-}}^{Total}$.BR($W^{-}\rightarrow l^{-}\nu)$ at $\sqrt{s}$=7, 8 and 13 TeV.}
\label{fig:7813}
\end{center}
\end{figure}

Figure~\ref{fig:7813} illustrates the comparison of measured total cross sections of $W^{+}$ (left) and $W^{-}$ (right) \cite{Cms7,Cms8,Cms13} with theoretical predictions at $\sqrt{s}$= 7, 8, and 13 TeV. Here, the green band shows Compact Muon selenoid (CMS) group results including only systematic uncertainty while the yellow band presents the CMS results with the quadrature sum of the reported systematic and luminosity uncertainties. On the comparison plot, it is clearly observed that HERAPDF2.0 provides the best description of measured $W^{\pm}$ cross section for all reported collision energies and its prediction is higher than other three PDF models. Because of this, HERAPDF2.0 is chosen as a reference prediction after this point in this paper. \par
\begin{figure*}[h!]
\begin{center}
\includegraphics[width=0.40\textwidth,height=0.16\textheight]{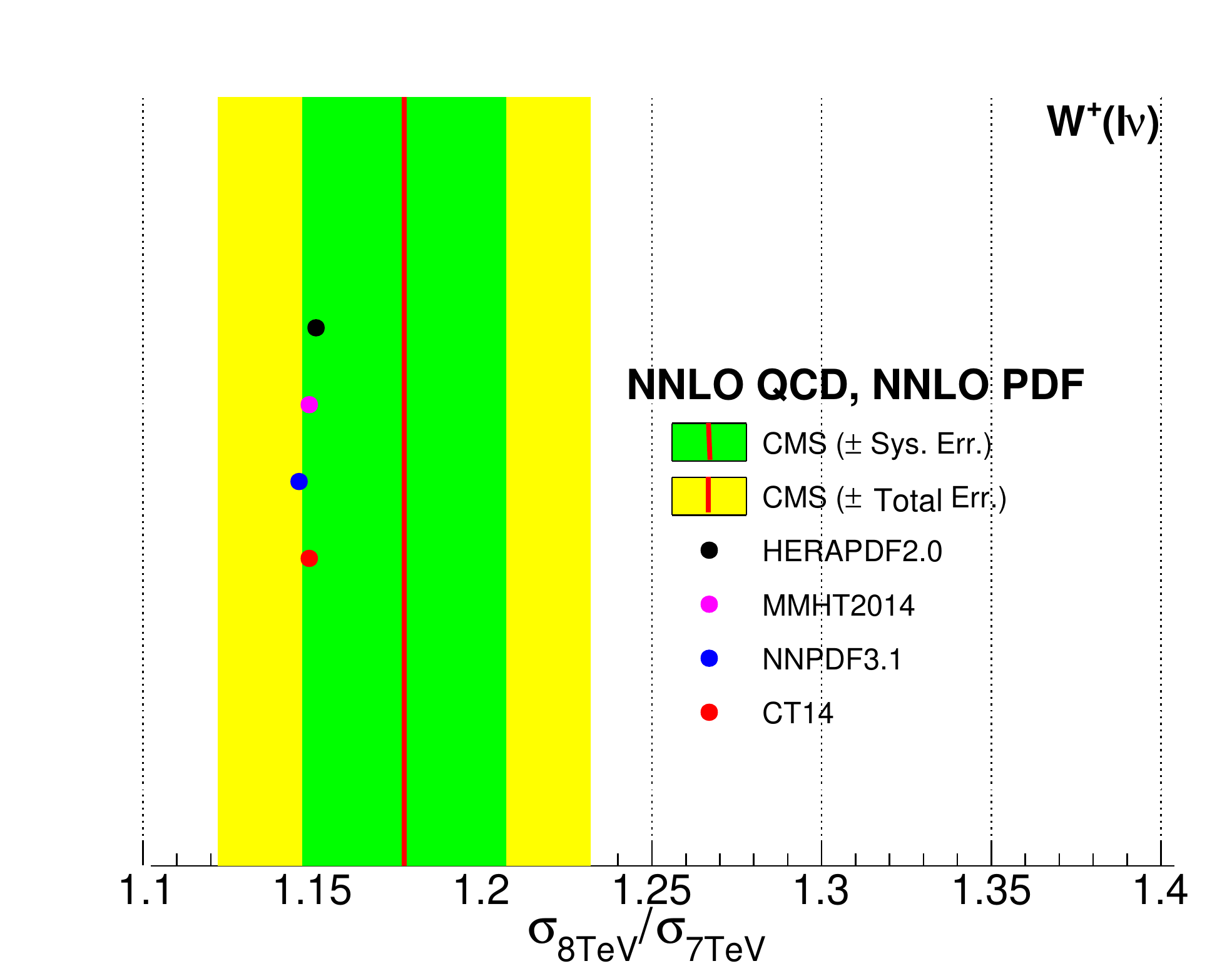}
\includegraphics[width=0.40\textwidth,height=0.16\textheight]{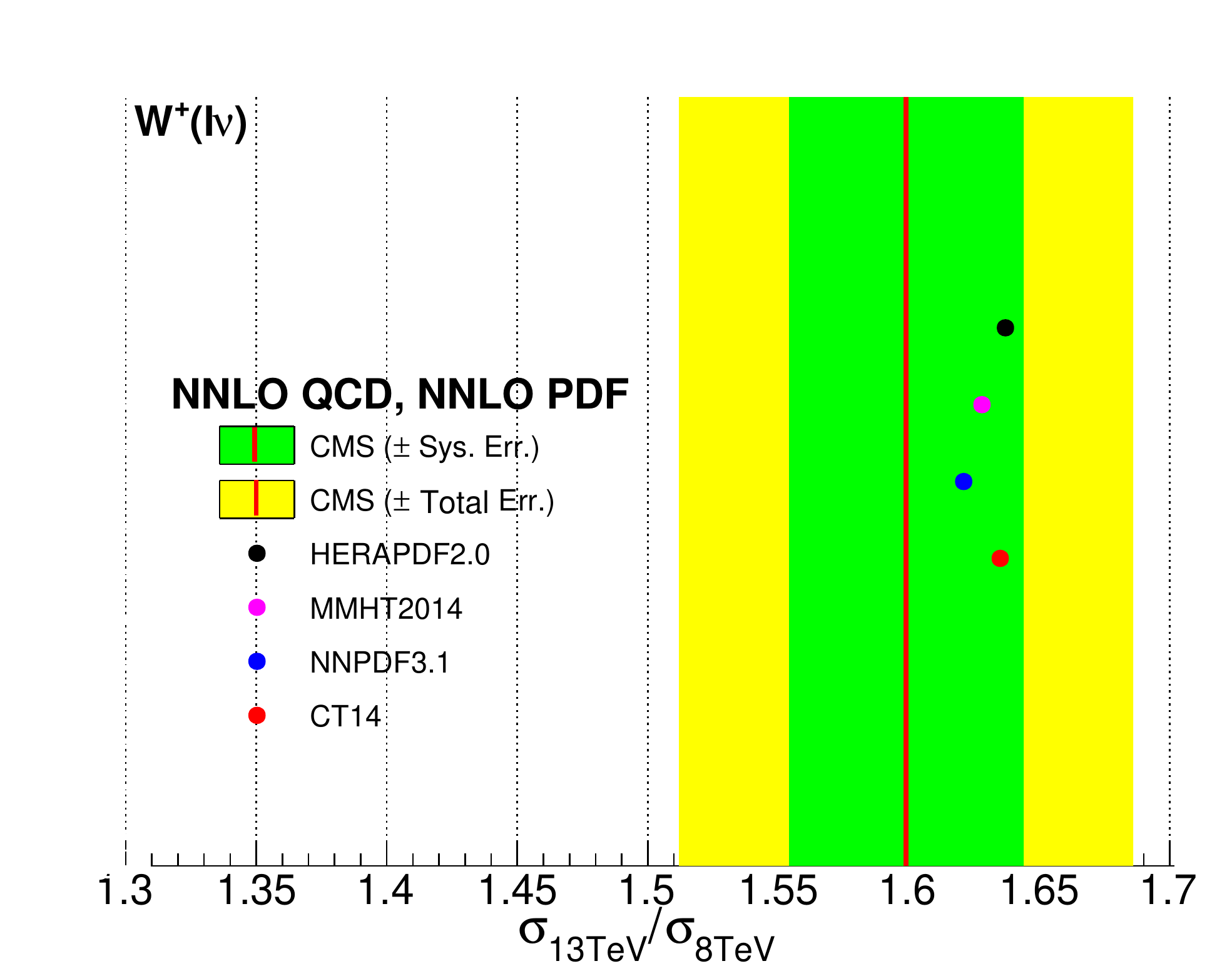}
\includegraphics[width=0.40\textwidth,height=0.16\textheight]{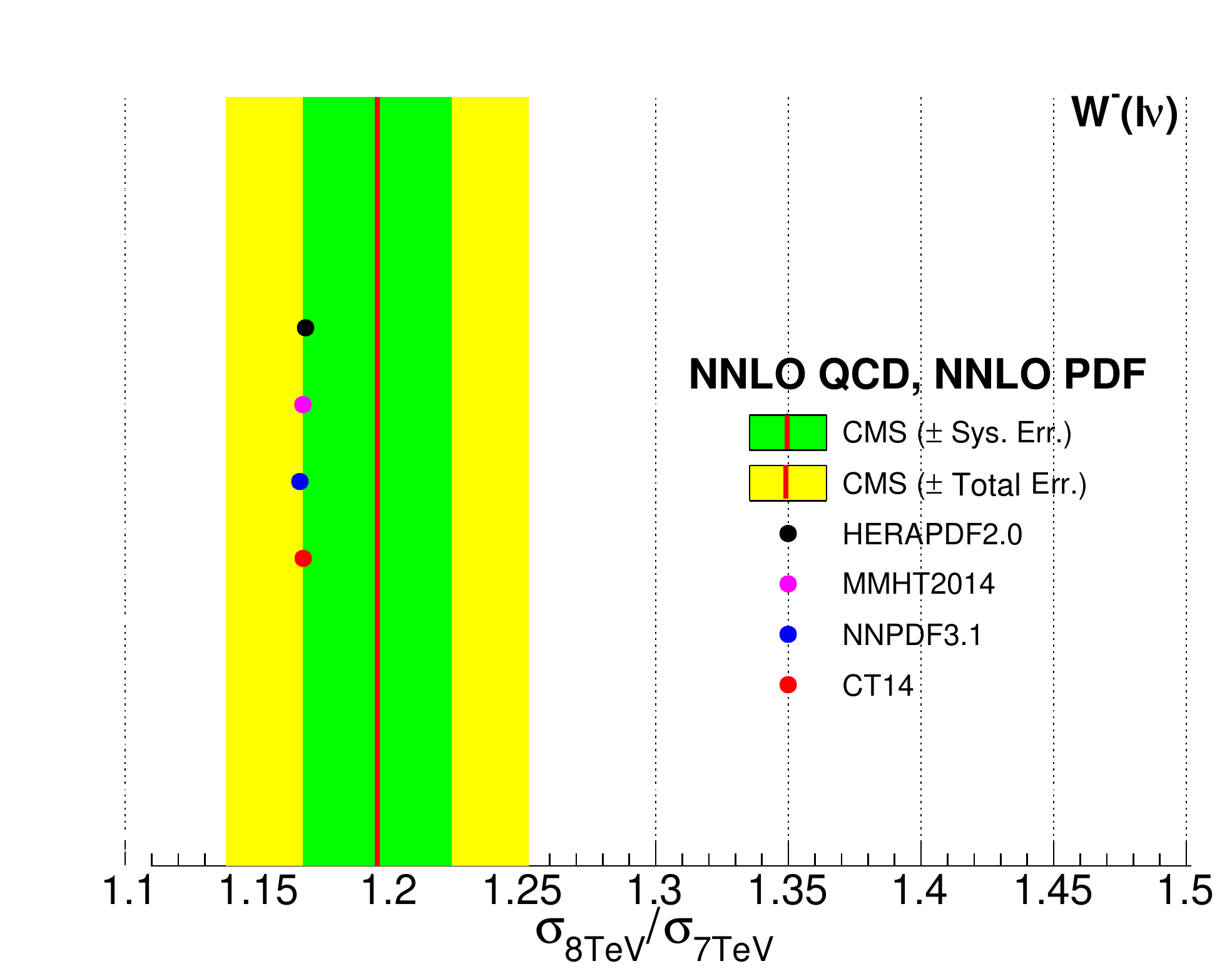}
\includegraphics[width=0.40\textwidth,height=0.16\textheight]{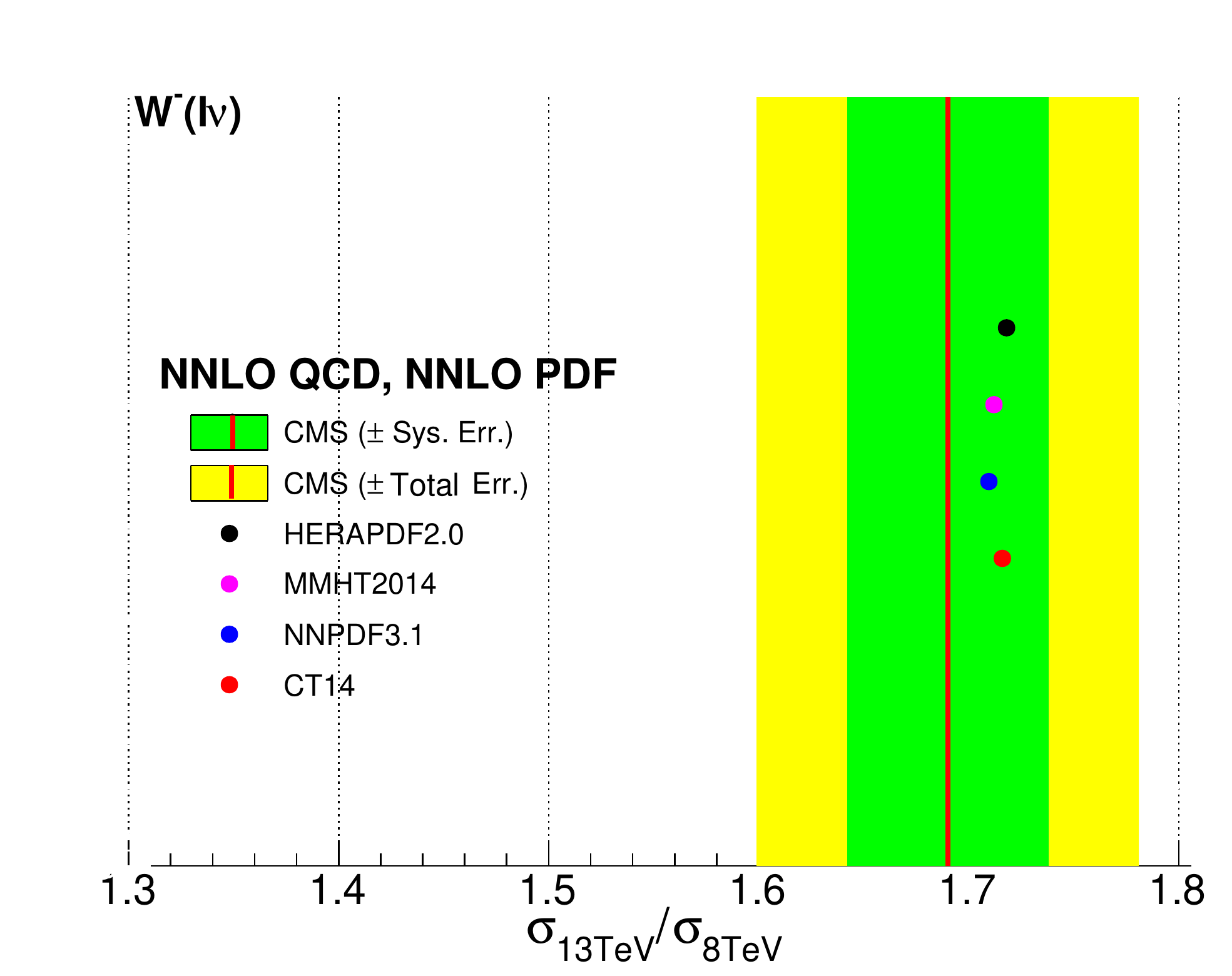}
\caption{Predicted and measured total cross section ratios of $\sigma_{W^{\pm}}^{8TeV}/\sigma_{W^{\pm}}^{7TeV}$ and $\sigma_{W^{\pm}}^{13TeV}/\sigma_{W^{\pm}}^{8TeV}$. The upper and lower rows are for $W^{+}$ and $W^{-}$, respectively. }
\label{fig:comp7813}
\end{center}
\end{figure*}

The left column of Figure~\ref{fig:comp7813} indicates the comparison of the measured total cross section ratios $\sigma_{8TeV}^{W^{+}}/\sigma_{7TeV}^{W^{+}}$ (upper) and $\sigma_{8TeV}^{W^{-}}/\sigma_{7TeV}^{W^{-}}$ (lower) with theoretical predictions. Similarly, the right column of Figure~\ref{fig:comp7813} shows the comparison of the measured $\sigma_{13TeV}^{W^{+}}/\sigma_{8TeV}^{W^{+}}$ (upper) and $\sigma_{13TeV}^{W^{-}}/\sigma_{8TeV}^{W^{-}}$ (lower) with the corresponding predictions. Here, the green band stands for the CMS result with only systematic uncertainty while the yellow band presents the CMS result with the quadrature sum of the reported systematic and luminosity uncertainties. The red line on the figures indicates the ratios of central values. Here, it is found that the considered PDF models provide the predictions within the systematic uncertainty band of the experimental results, and all ratios are very close to each other.\par
\begin{figure}[h!]
\begin{center}
\includegraphics[width=0.4\textwidth,height=0.2\textheight]{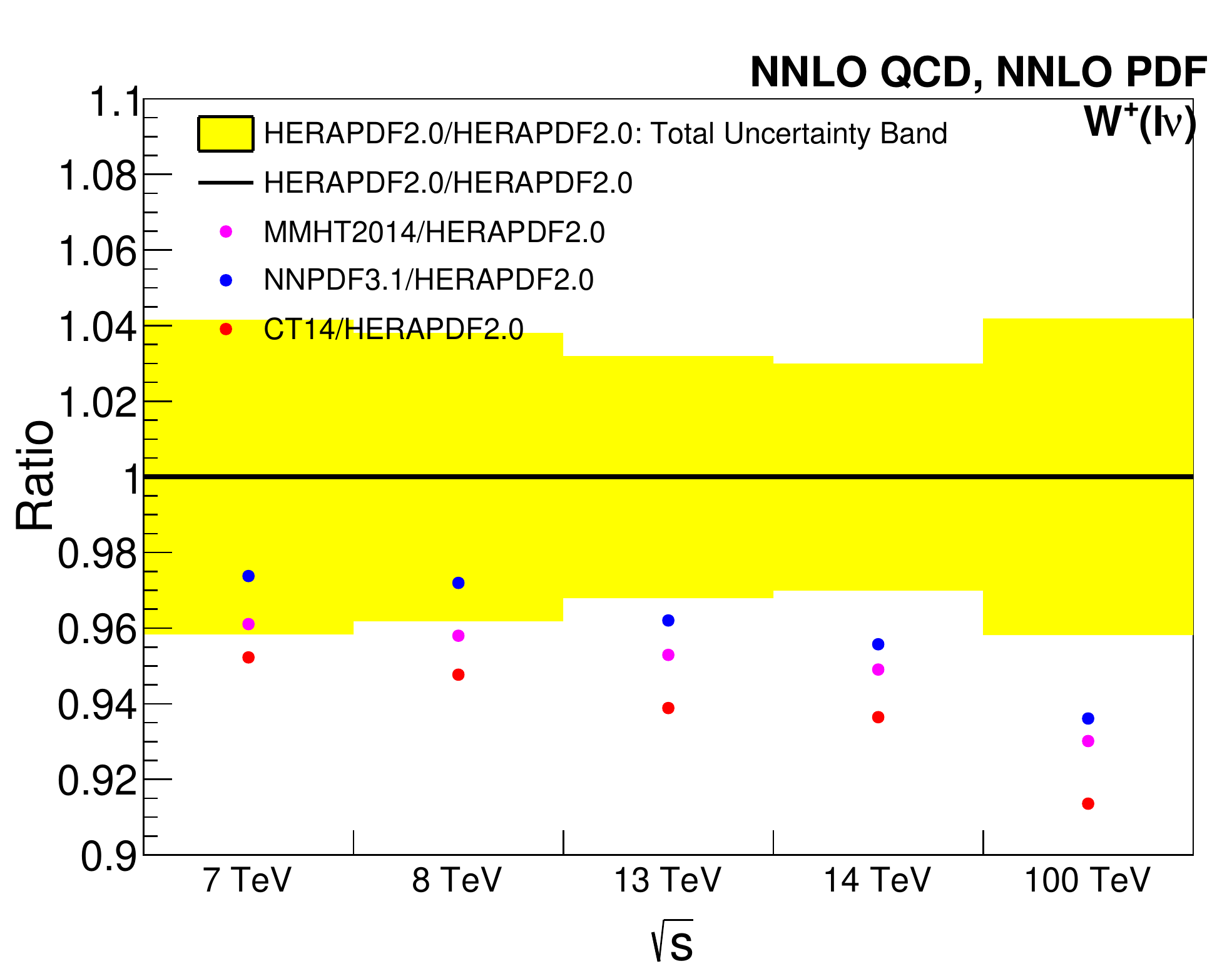}
\includegraphics[width=0.4\textwidth,height=0.2\textheight]{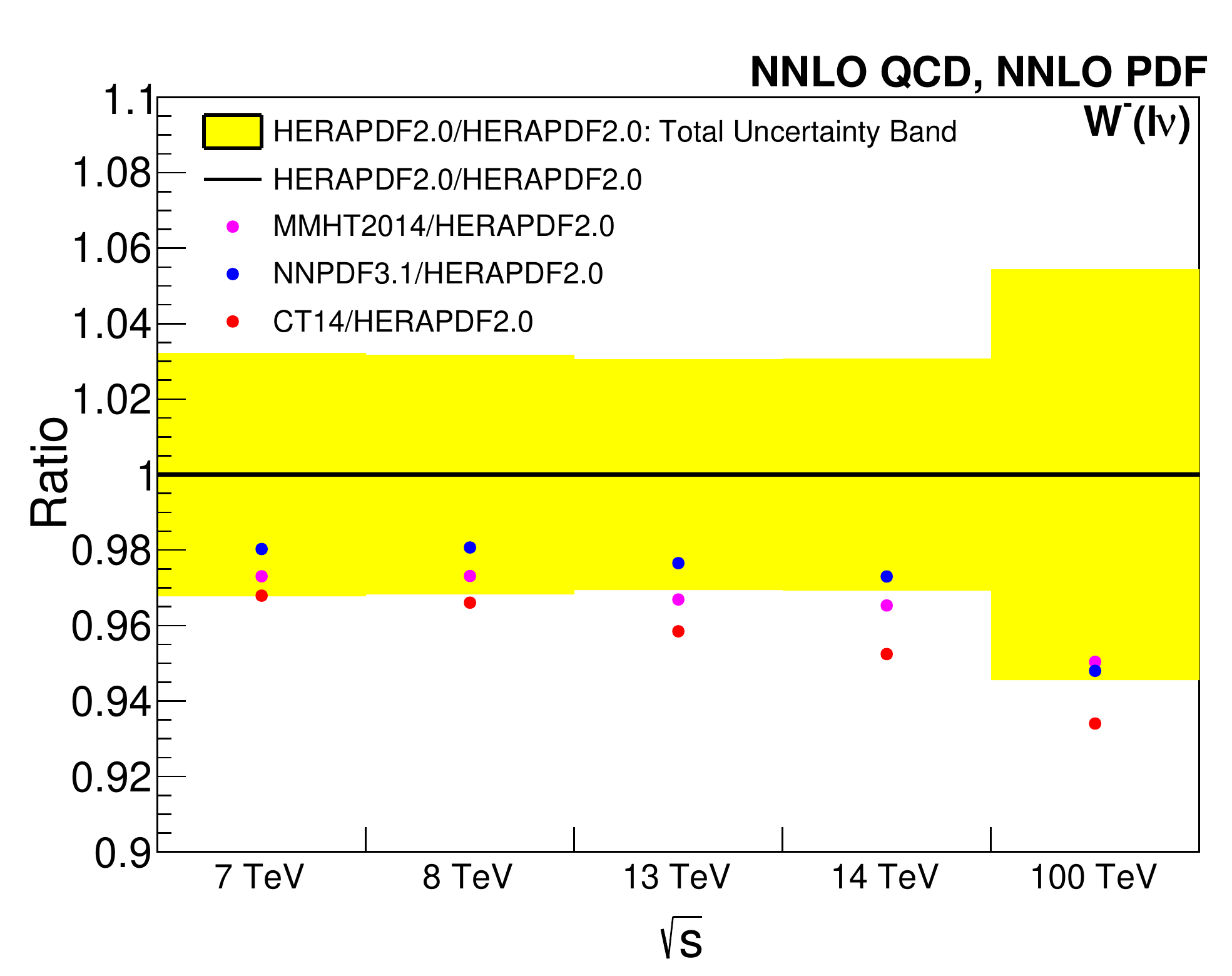}
\caption{Comparison of HERAPDF2.0 prediction with MMHT2014, NNPDF3.1 and CT14 predictions at 7, 8, 13, 14, and 100 TeV in the fiducial region. The upper and lower figures show the results of $W^{+}$ and $W^{-}$, respectively.}
\label{fig:comp781314100}
\end{center}
\end{figure}

The fiducial cross sections are further computed for 7, 8, 13, 14, and 100 TeV proton-proton ($pp$) collisions. HERAPDF2.0 NNLO QCD prediction is taken as reference, and Figure~\ref{fig:comp781314100} is created to make comparisons of reference distribution with other considered PDF models. The upper and lower plots of Figure~\ref{fig:comp781314100} illustrate these comparisons for $W^{+}$ and $W^{-}$, respectively. Here, the yellow band stands for total uncertainty of HERAPDF2.0 prediction. Especially, the differences between HERAPDF2.0 and other PDF models are greater for fiducial cross section of $W^{+}$ than what are observed for $W^{-}$. Beside, the biggest difference is with CT14 but CT14, MMHT2014, and NNPDF3.1 provide closer results to each other. Another remark should be underlined here is that the disagreement between HERAPDF2.0 and other three increases by the increase of the collision energy, and this is true for both $W^{+}$ and $W^{-}$ bosons.\par
Since there is an obvious difference between HERAPDF2.0 and other considered PDF (MMHT2014, NNPDF- 3.1, and CT14) predictions in the fiducial region, correlation ellipses are also drawn. Figure~\ref{fig:comp781314100} shows the predicted $W^{\pm}$ at 13 TeV versus  $W^{\pm}$ at 8 TeV boson and $W^{\pm}$ at 100 TeV versus $W^{\pm}$ at 14 TeV boson production cross sections times branching fractions. The ellipses illustrate the 68$\%$ CL coverage for only PDF uncertainties with solid lines. Since all considered uncertainties are calculated only for HERAPDF2.0, an ellipse with dashed line is drawn for HERAPDF2.0 including the quadrature sum of the PDF, scale, $\alpha_{s}$, model, and parameterization errors. It can be safely concluded that MMHT2014, NNPDF3.1 and CT14 predictions are in well agreement but HERAPDF2.0 provides higher cross section values than other three. \par
Beside all of these, the collision data recorded at higher collision energy provides more number of events for particles than data recorded at lower collision energy if the recorded data amount is same at higher and lower collisions. Because of this, it is advantageous to increase collision energy at hadron colliders. This statement could be more clear if numerical numbers are provided: the number of events recorded in a $pp$ collision could be written as follows:
\begin{center}
Number of events=$\sigma$L
\end{center}
\noindent where $\sigma$ is the cross section and L is the luminosity of data. Considering the same number of events at both 13 and 14 TeV, the following equation could be set:
\begin{center}
$\sigma_{13}L_{13}=\sigma_{14}L_{14}\rightarrow L_{14}=\frac{\sigma_{13}}{\sigma_{14}}L_{13}$\\
\end{center}
This equation shows the cross section relationship between 13 and 14 TeV. Here, the total recorded data amount at 13 TeV by the CMS detector is 84.76 fb$^{-1}$ by November 2017 as reported in ref~\cite{cmslumi}. To satisfy the equation, the cross section ratios are calculated using the FEWZ 3.1 interfaced with NNLO HERAPDF2.0 PDF model at NNLO QCD.
\begin{center}
$L_{14}^{W^{+}}=\frac{\sigma_{13}^{W^{+}}}{\sigma_{14}^{W^{+}}}L_{13}$=0.941x84.76= 79.76 fb$^{-1}$\\
$L_{14}^{W^{-}}=\frac{\sigma_{13}^{W^{-}}}{\sigma_{14}^{W^{-}}}L_{13}$=0.9286x84.76= 78.71 fb$^{-1}$\\
\end{center}
As can be seen from the calculations if 79.76 fb$^{-1}$ data at 14 TeV were recorded by the CMS, the same statistics of $W^{+}$ and $W^{-}$ with 13 TeV data would be reached. The same calculation is done for 100 TeV collider below:
\begin{center}
$\sigma_{13}L_{13}=\sigma_{100}L_{100}\rightarrow L_{100}=\frac{\sigma_{13}}{\sigma_{100}}L_{13}$\\
$L_{100}^{W^{+}}=\frac{\sigma_{13}^{W^{+}}}{\sigma_{100}^{W^{+}}}L_{13}$=0.2126x84.76=18.02 fb$^{-1}$\\
$L_{100}^{W^{-}}=\frac{\sigma_{13}^{W^{-}}}{\sigma_{100}^{W^{-}}}L_{13}$=0.1665x84.76=14.11 fb$^{-1}$\\
\end{center}
This time, one could reach to 13 TeV $W^{+}$ and $W^{-}$ statistics with 15.18 fb$^{-1}$ recorded data at 100 TeV.  These calculations prove that higher number of events for particles could be obtained at higher collision energies if the same amounts of data were recorded.
\subsection{Uncertainties}
\begin{figure}[h!]
\begin{center}
\includegraphics[width=0.45\textwidth,height=0.18\textheight]{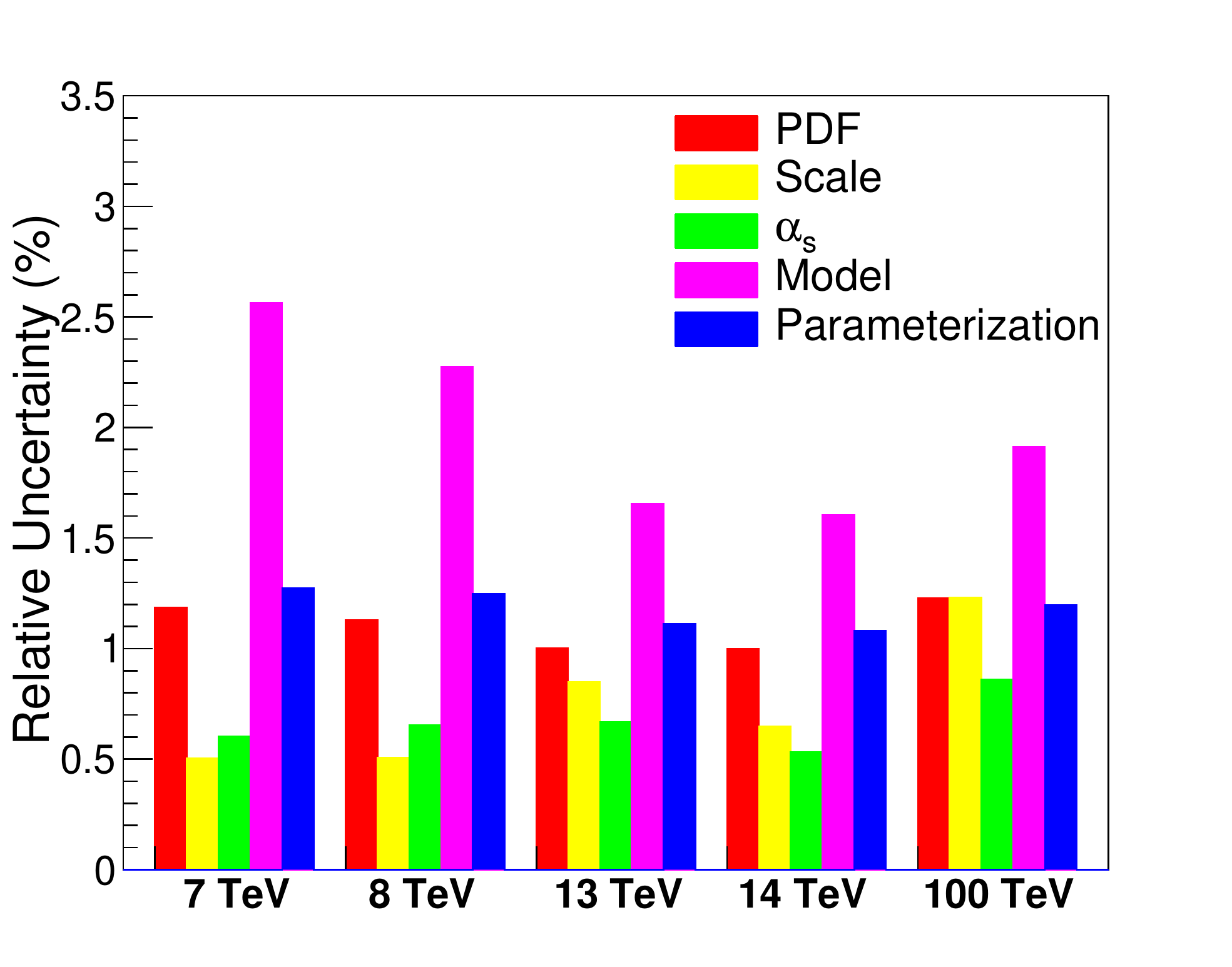}
\includegraphics[width=0.45\textwidth,height=0.18\textheight]{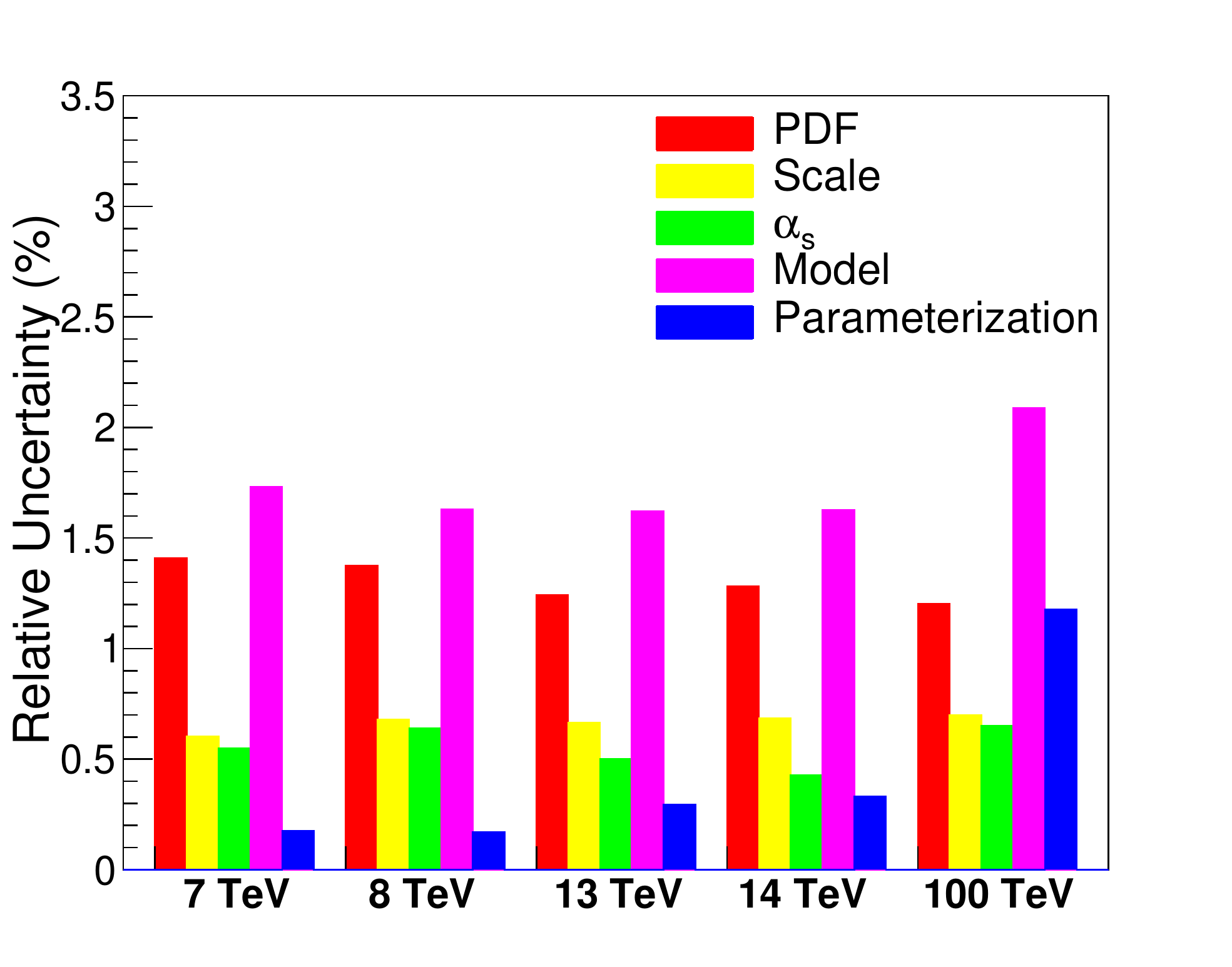}
\caption{The changes on the uncertainties by the increase of the collision energy at NNLO QCD. The upper and lower figures present the variations for $W^{+}$ and $W^{-}$, respectively.}
\label{fig:wpmunc}
\end{center}
\end{figure}

In this section, PDF, scale, $\alpha_{s}$, model, and parameterization uncertainties are considered and calculated only for the reference PDF model (HERAPDF2.0) at NNLO QCD. This lets us to discuss the variation of each uncertainty by the increase of the collision energy. The PDF uncertainty is estimated following closely the prescription of PDF4LHC working group \cite{PDFLHC,PDFLHC2}. The scale uncertainties are calculated varying the values of $\mu_{R}$ and $\mu_{F}$ scales by factors of two, $M_{W}/2\leq\mu_{R}\leq2M_{W}$ and $M_{W}/2\leq\mu_{F}\leq2M_{W}$ where $\mu_{R}$ and $\mu_{F}$ are varied independently from each other. The default value of $\mu_{R,F}$ is set to $M_{W}$, which is taken as 80.304 GeV. The maximum value of the variation is taken as the scale uncertainty, which makes the scale uncertainty symmetric. The $\alpha_{s}$ uncertainty is estimated varying $\alpha_{s}$ by 0.001. Then, the uncertainties are symmetrized by taking the bigger value from estimated up and down errors. To calculate the parameterization and model uncertainties, HERAPDF20$\_$NNLO$\_$VAR PDF set is used. The PDF set has 13 eigenvectors, and each eigenvector is treated one by one to estimate the model and parameterization errors. First, model and parameterization errors are calculated separately, and then, they are summed in quadrature. More details about the calculation of model and parameterization errors can be found in \cite{paramodel}. After the calculation of all considered uncertainties, the total uncertainty is determined by adding the individual uncertainties in quadrature. The numerical values are provided in Table~\ref{table:wpmunc} and illustrated in Figure~\ref{fig:wpmunc}.
\begin{table*}
\caption{Relative uncertainty of NNLO $W^{\pm}$ HERAPDF2.0 prediction at 7, 8, 13, 14, and 100 TeV. All numbers are stated as the percentage($\%$) of the central value of the prediction.}
\label{table:wpmunc}
\begin{tabular*}{\textwidth}{@{\extracolsep{\fill}}lccccc@{}}
\hline
&7 TeV & 8 TeV & 13 TeV & 14 TeV & 100 TeV\\
\hline
\multicolumn{6}{c}{Values for $W^{+}\rightarrow l^{+}\nu$} \\ 
$\sigma_{Fid}$ [pb]&3461.93&3852.04&5594.94&5946.55&26318\\
PDF ($\%$)&1.187& 1.128&1.002&0.999&1.227 \\ 
Scale ($\%$)& 0.504 &0.508 & 0.850&0.648&1.232\\
$\alpha_{s}$ ($\%$)& 0.603& 0.653&0.668&0.531&0.860\\
Model ($\%$)& 2.562& 2.276&1.655&1.603&1.913\\
Parameterization ($\%$)&1.274 &1.247 &1.111&1.080&1.197\\
Total Err. ($\%$) &2.945 &2.694&2.266&2.124&2.957\\
\hline
\multicolumn{6}{c}{Values for $W^{-}\rightarrow l^{-}\nu$} \\ 
$\sigma_{Fid}$ [pb]&2269.29&2594.6&4181.66&4503&25115.7\\
PDF ($\%$)& 1.410& 1.375&1.242&1.281&1.202 \\ 
Scale ($\%$)&0.603 & 0.679&0.666&0.685&0.699\\
$\alpha_{s}$ ($\%$)& 0.548 & 0.639&0.500&0.429&0.650\\
Model ($\%$)& 1.733& 1.630&1.620&1.626&2.089\\
Parameterization ($\%$)&0.177 & 0.171&0.296&0.331&1.176\\
Total Err. ($\%$)&2.285&2.245&2.160&2.158&3.848\\
\hline
\end{tabular*}
\end{table*}
The table and figure are well indicative of the uncertainty change by the increase of collision energy. The total uncertainty is obtained by adding all considered uncertainties in quadrature, and it is generally greater for $W^{+}$ than the one calculated for $W^{-}$ boson. It is also found that the total uncertainty decreases by the increase of the collision energy with only one exception that it increases while going from 14 to 100 TeV. Another finding should be highlighted here is that model uncertainty is the greatest uncertainty for both $W^{+}$ and $W^{-}$ at all considered center-of-mass energies, which is followed by PDF uncertainty.

\subsection{W boson lepton charge asymmetry}
\begin{figure}[h!]
\begin{center}
\includegraphics[width=0.4\textwidth,height=0.2\textheight]{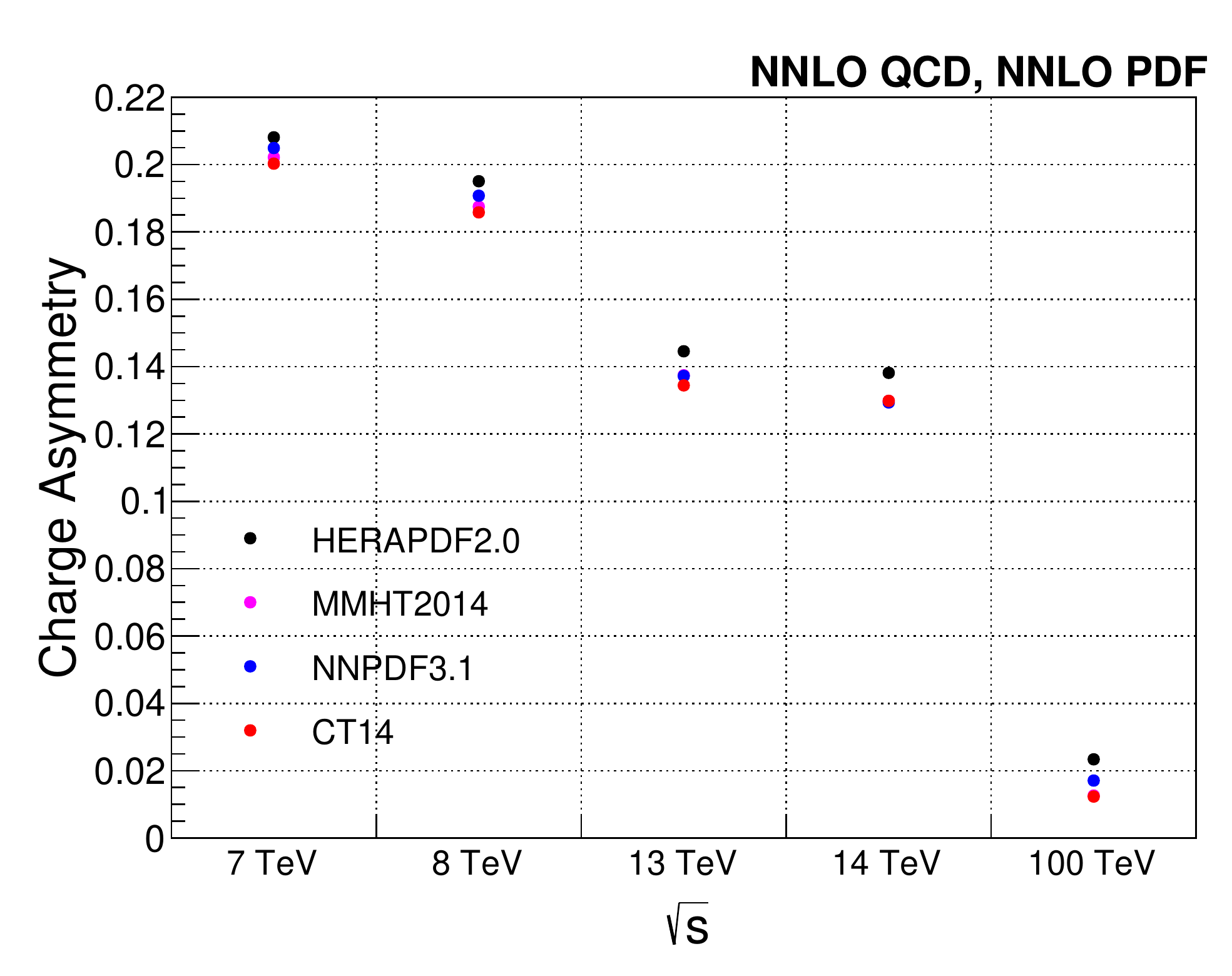}
\includegraphics[width=0.4\textwidth,height=0.2\textheight]{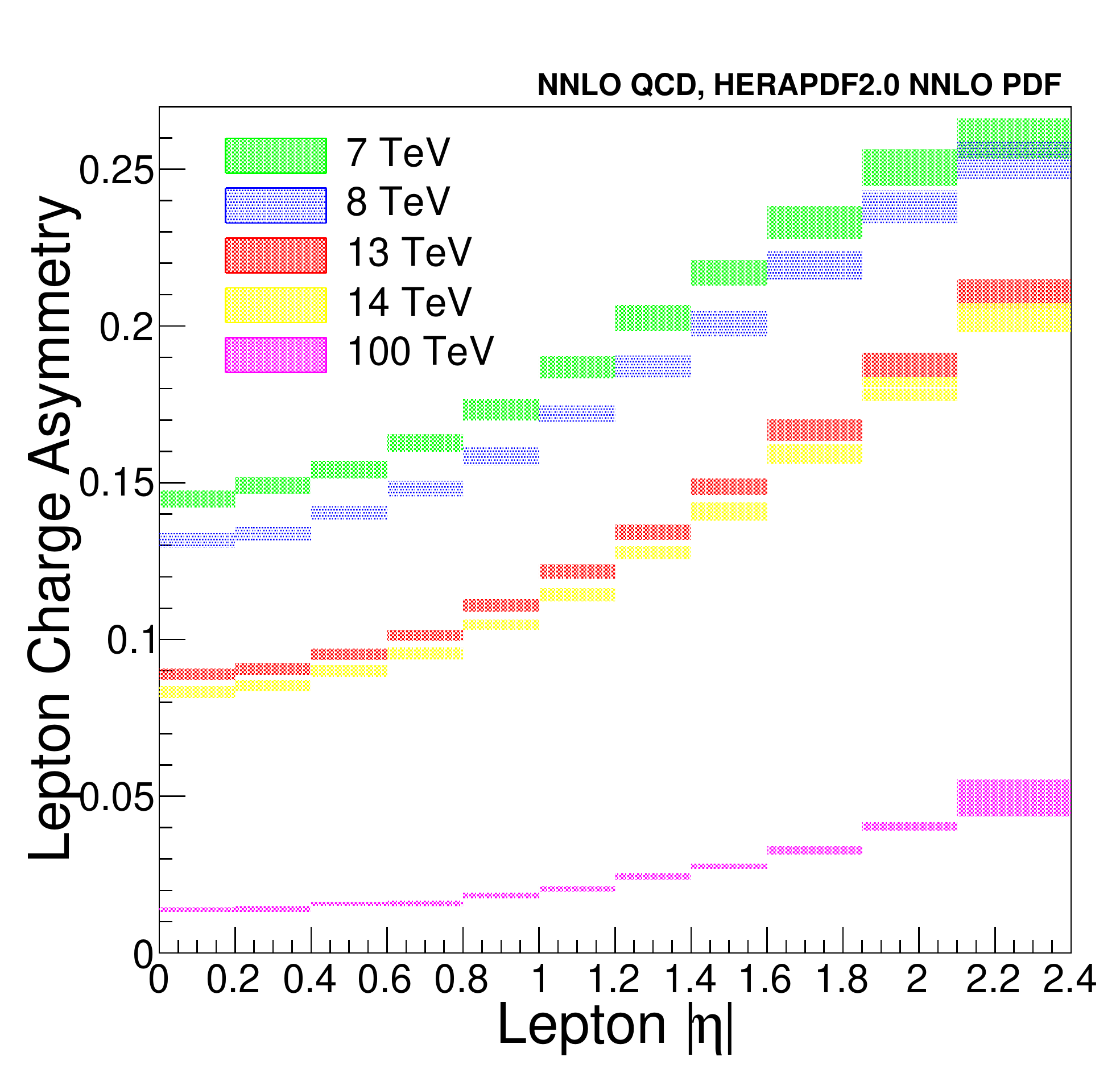}
\caption{W boson charge asymmetry. The upper plot shows the charge asymmetry as a function of collision energy, and the lower figure illustrates the charge asymmetry as a function of absolute lepton pseudorapidity.}
\label{fig:wasym}
\end{center}
\end{figure}

A detailed description and discussion on W boson charge asymmetry can be found in \cite{hasanQCD,hasanturk}. The asymmetry depends on d/u ratio and the difference between d and u quarks gets smaller by going higher center-of-mass energies \cite{hasanQCD,hasanturk}. It is therefore expected a decrease of the asymmetry by the increase of the collision energy. Simply, the definition of W boson charge asymmetry is given as follows:
\begin{equation} \label{eq:asymm1}
A_{W}=\frac{\sigma(W^{+}\rightarrow l^{+}\nu) -\sigma(W^{-}\rightarrow l^{-}\nu)}{\sigma(W^{+}\rightarrow l^{+}\nu) + \sigma(W^{-}\rightarrow l^{-}\nu) }
\end{equation}
Experimentally, more directly accessible quantity is the lepton charge asymmetry as a function of the charged lepton pseudorapidity ($\eta$) that encompasses the beam angle:
\begin{equation} \label{eq:asymm2}
A_{W}(\eta)=\frac{ \frac{d\sigma}{d\eta}(W^{+}\rightarrow l^{+}\nu) - \frac{d\sigma}{d\eta}(W^{-}\rightarrow l^{-}\nu)}{ \frac{d\sigma}{d\eta}(W^{+}\rightarrow l^{+}\nu) + \frac{d\sigma}{d\eta}(W^{-}\rightarrow l^{-}\nu) }
\end{equation}
\noindent where $d\sigma/d\eta$ is the differential cross section of W boson production. Based on these equations, the charge asymmetries are calculated in the fiducial region and illustrated in Figure~\ref{fig:wasym}. The upper plot presents the result of Equation~\ref{eq:asymm1} while the lower figure indicates the result of Equation~\ref{eq:asymm2}. Here, the upper plot shows that the considered PDF models are in well agreement. The lower plot is produced using NNLO HERAPDF2.0 model at NNLO QCD. The colorful bands show only PDF uncertainty of HERAPDF2.0 for each energy level. As it is expected, the asymmetry decreases by the increase of the collision energy. 

\section{Results and Discussion}
In this paper, some properties of $W^{\pm}$ are studied using FEWZ3.1 program interfaced with the latest modern PDF models to provide an insight to transition from the LHC to 100 TeV collider. The LHC runs are already recorded at 7, 8, and 13 TeV, and 14 TeV collisions are expected to be happening after the intended shutdown in 2019. 14 TeV collisions will be the last milestone of the LHC since it reaches its maximum capacity. 100 TeV $pp$ collider is going to be useful to make new discoveries and to investigate the properties of existing particles. Therefore, $W^{\pm}$ boson rates, the uncertainties of PDF, scale, $\alpha_{s}$, model, and parameterization for $W^{\pm}$, and W boson lepton charge asymmetry are examined by varying the collision energy from 7 to 100 TeV.\par

HERAPDF2.0 is found to be the best describer of experimental results at 7, 8, and 13 TeV for $W^{\pm}$; therefore, it is chosen as the reference PDF model for the comparisons. The difference between the reference prediction and others increases by the increase of the collision energy. In terms of uncertainty comparisons of W, the biggest uncertainty source is the model uncertainty at all energy levels, which is followed by PDF uncertainty. W boson charge asymmetry (as a function of lepton pseudorapidity) are further calculated. By the increase of the collision energy from 7 to 100 TeV, the asymmetry decreases and gets closer to zero as expected. FEWZ3.1 predictions are found to have remarkable agreement with the CMS measurement. \par


\begin{thebibliography}{}
\bibitem{atlashiggs} ATLAS Collaboration, Phys. Lett. B {\bf 716}, 1-29 (2012).
\bibitem{cmshiggs} CMS Collaboration, Phys. Lett. B {\bf 716}, 30-61 (2012).
\bibitem{LHC} WEB: http://home.cern/topics/large-hadron-collider, 15 September 2017. 
\bibitem{SM} M. J. Herrero, {\em The Standard Model}, arXiv:hep-ph/9812242.
\bibitem{FEWZ} Y. Li and F. Petriello, Phys. Rev. D {\bf 86}, 094034 (2012).
\bibitem{hasanQCD} H. Ogul {\it et al.}, Adv. High Energy Phys. {\bf 2016}, 7865689 (2016).
\bibitem{hasanZboson} H. Ogul and K. Dilsiz, Adv. High Energy Phys. {\bf 2017}, 8262018 (2017).
\bibitem{chargeasymmetry7} CMS Collaboration, Phys. Rev. D {\bf 90}, 032004 (2014).
\bibitem{chargeasymmetry8} CMS Collaboration, Eur. Phys. J. C {\bf 76}, 469 (2016). 
\bibitem{herapdf20} H. Abramowicz {\it et al.}, Eur. Phys. J. C {\bf 75}, 580 (2015).
\bibitem{mmht2014} L. A. Harland-Lang, A. D. Martin, P. Motylinski, R. S. Thorne, Eur. Phys. J. C {\bf 75}, 204 (2015).
\bibitem{nnpdf31} NNPDF Collaboration, {\em Parton distributions from high-precision collider data}, arXiv:1706.00428.
\bibitem{ct14pdf} S. Dulat {\it et al.}, Phys. Rev. D {\bf 93}, 033006 (2016).
\bibitem{Cms7} CMS Collaboration, J. High Energ. Phys. \textbf{10}, 132 (2011).
\bibitem{Cms8} CMS Collaboration, Phys. Rev. Lett. \textbf{112}, 191802 (2014). 
\bibitem{Cms13} CMS Collaboration, "Inclusive W/Z cross section at 13 TeV", SMP-15-004 (2015).
\bibitem{cmslumi} WEB: https://twiki.cern.ch/twiki/bin/view/CMSPublic/ LumiPublicResults, 10 November 2017
\bibitem{PDFLHC} J. Butterworth {\it et al.}, J. Phys. G. {\bf 43}, 023001 (2016).
\bibitem{PDFLHC2} A. Buckley {\it et al.},  Eur. Phys. J. C {\bf 75}, 132 (2015).
\bibitem{paramodel} H1 and ZEUS Collaborations, Eur. Phys. J. C {\bf 75}, 580 (2015).
\bibitem{hasanturk} H. Ogul, Turk J Phys. {\bf 41/5}, 520 (2017).
\bibitem{kfact1} R. Vogt, Heavy Ion Physics {\bf 17/1}, 75-92 (2003).
\bibitem{kfact2} G. Davatz {\it et al.}, J. High Energ. Phys. {\bf 2004}, JHEP05 (2004). 
\bibitem{cmsz13} CMS Collaboration, ``Measurement of inclusive and differential Z boson production cross sections in $pp$ collisions at $\sqrt{s}=13$ TeV", CMS-PAS-SMP-15-011 (2016).
\bibitem{afbcms8} CMS Collaboration, Eur. Phys. J. C {\bf 76}, 325 (2016).
\end{thebibliography}
\end{document}